\documentclass[preprint,showpacs,showkeys,preprintnumbers,amsmath,amssymb,aip,apl]{revtex4-1}

\usepackage{graphicx}
\usepackage{float}
\usepackage{color}
\begin{document}
\preprint{MnCoGa thin films}

\title{Perpendicularly magnetized Mn-Co-Ga-based thin films with high coercive field.}

\author{Siham Ouardi}
\affiliation{Max Planck Institute for Chemical Physics of Solids,
             01187 Dresden, Germany.}

\author{Takahide Kubota}
\affiliation{WPI-Advanced Institute for Materials Research (WPI-AIMR), Tohoku University, Sendai 980-8577, Japan}
            
\author{Gerhard H. Fecher}
\affiliation{Max Planck Institute for Chemical Physics of Solids,
             01187 Dresden, Germany.}
             
\author{Rolf Stinshoff}
\affiliation{Max Planck Institute for Chemical Physics of Solids,
             01187 Dresden, Germany.}
 
\author{Shigemi Mizukami}
\affiliation{WPI-Advanced Institute for Materials Research (WPI-AIMR), Tohoku University, Sendai 980-8577, Japan}

\author{Terunobu Miyazaki}
\affiliation{WPI-Advanced Institute for Materials Research (WPI-AIMR), Tohoku University, Sendai 980-8577, Japan}

\author{Eiji Ikenaga}
\affiliation{Japan Synchrotron Radiation Research Institute, SPring-8, Hyogo 679-5198, Japan}
            
\author{Claudia Felser}
\email{felser@cpfs.mpg.de}
\affiliation{Max Planck Institute for Chemical Physics of Solids,
             01187 Dresden, Germany.}

\date{\today}

\begin{abstract}

Mn$_{3-x}$Co$_{x}$Ga epitaxial thin films were grown on MgO substrates by magnetron 
co-sputtering. Structures were tetragonal or cubic depending on Co content.
Composition dependence of saturation 
magnetization and uniaxial magnetic anisotropy $K_u$ of the films were investigated. 
A high $K_u$ (1.2~MJ m$^{-3}$) was achieved for the Mn$_{2.6}$Co$_{0.3}$Ga$_{1.1}$ film 
with the magnetic moment 0.84$\mu_B$. Valence band spectra were obtained by
 hard X-ray photoelectron spectroscopy. Sharp peaks in the cubic case, which were absent in the 
tetragonal case, prove that a {\it van~Hove} singularity causes 
a {\it band Jahn--Teller effect} with tetragonal distortion. Observations agree well with 
the first--principles calculations.

\end{abstract}

%\pacs{71.20.Lp, 79.60.-i}

\maketitle

%%%%%%%%%%%%%%%%%%%%%%%%%%%%%%%%%%%%%%%%%%%%%%%%%%%%%%%%%%%%%%%%%%%%%%
%\section{Introduction}

The tetragonally distorted Heusler compound Mn$_3$Ga with a DO$_{22}$ structure
has attracted considerable attention as a potential candidate for spin-transfer torque (STT) applications. 
Mn$_3$Ga has been proposed for use in STT applications as a compensated ferrimagnet 
exhibiting half-metallicity with 88\% polarization, 
high Curie temperature T$_c$ (730~K), and hard magnetic properties ~\cite{BFW07,WBF08}. 
The epitaxial growth of Mn-Ga thin films with giant perpendicular magnetic anisotropy (PMA)~\cite{WMW09,KRV11,MWS11,MKW12} 
and high tunnel magnetoresistance (TMR)~\cite{KAM11,KMM12} have been successfully realized. 
According to the Slonczewski - Berger equation~\cite{Slo96,Be96}, in order to optimize materials for
STT application (such as minimization of the switching current), the saturation magnetization $M_s$ and 
Gilbert damping have to be minimized. The use of off-stoichiometric Mn$_3$Ga with Mn deficit demonstrates the 
possibility of increasing the magnetic energy products of the material; 
however, $M_s$ also correspondingly increases~\cite{WBF08}.

Recently, it has been shown that partial  substitution of Mn by Co in Mn$_{3-x}$Co$_x$Ga leads to a 
reduced saturation magnetization $M_s$~\cite{AWF11,WCG12}. 
The system exhibits a tetragonal structure for Co concentrations lower than $x = 0.4$. 
Similar to Mn$_3$Ga, it crystallizes in a tetragonally distorted variation of the Heusler 
structure and exhibits comparably hard ferrimagnetic properties. On the other hand, Co-rich alloys ($x>0.5$) 
crystallize as cubic and magnetically soft structures. 
The tetragonal distortion of the cubic Heusler structure is caused by  
electronic instabilities corresponding to a band--type Jahn–-Teller effect~\cite{CKF12}. 
However, in contrast to the Mn$_{3-x}$Rh$_x$Sn system, the Curie temperature is still high~\cite{WCG12}.
Magnetic circular dichroism in X-ray absorption (XMCD) has been used to confirm the ferrimagnetic 
character of the Mn$_{3-x}$Co$_x$Ga system, with Mn atoms occupying two different sublattices with antiparallel spin 
orientation and different degrees of spin localization~\cite{KJA11}. Ferrimagnetic characteristics were
 also similarly confirmed for epitaxial cubic thin films of Mn$_2$CoGa~\cite{MSK11}.
To identify the existence of a high density of states (DOS) at the 
Fermi energy ({\it van~Hove} singularity) in such films, a convenient experiment such as 
photoelecton spectroscopy is required to be performed.

In the present study, epitaxial thin films of Mn$_{3-x}$Co$_x$Ga with varying 
levels of Co content were grown directly on MgO substrates. 
The electronic structure of the films was determined by all-electron \textit{ab initio} calculations. The 
valence states of the films close to the Fermi energy were investigated by means of hard X-ray 
photoelectron spectroscopy, and the obtained results were compared with the calculations.
The magnetic properties of the films were investigated and compared with those of the corresponding bulk material.

%%%%%%%%%%%%%%%%%%%%%%%%%%%%%%%%%%%%%%%%%%%%%%%%%%%%%%%%%%%%%%%%%%%%%%
%\section{Experimental details}

Epitaxial 30-nm thin films with nominal compositions of Mn$_{2.6}$Co$_{0.3}$Ga$_{1.1}$ and 
Mn$_{2.1}$CoGa$_{0.9}$ were grown on a MgO (001) single crystalline substrate using an 
ultrahigh-vacuum magnetron sputtering system. A Mn-Ga target and an elemental Co 
target were used for the deposition. Structural analysis was performed by means of 
out-of-plane and in-plane X-ray diffractometers (XRD). The magnetic properties 
of the films were investigated using a vibrating sample magnetometer (VSM) with 
a maximum applied field of 2~T at room temperature.

The valence band spectra of the films were measured by hard X-ray 
photoelectron spectroscopy (HAXPES) using the undulator beamline BL47XU at SPring-8 
(Japan). Details of the HAXPES experiment have been previously reported~\cite{OSF11,OFK11,SOF12}. 
For comparison, we use the electronic structure calculations 
performed by means of the fully relativistic spin--polarized KKR (Korringa--Kohn--Rostoker) method
including a coherent potential approximation to account for random site occupation~\cite{EKM11}.

%%%%%%%%%%%%%%%%%%%%%%%%%%%%%%%%%%%%%%%%%%%%%%%%%%%%%%%%%%%%%%%%%%%%%%
%\section{Results and Discussion}

\begin{figure}[htb]
  \centering
  \includegraphics[width=8cm]{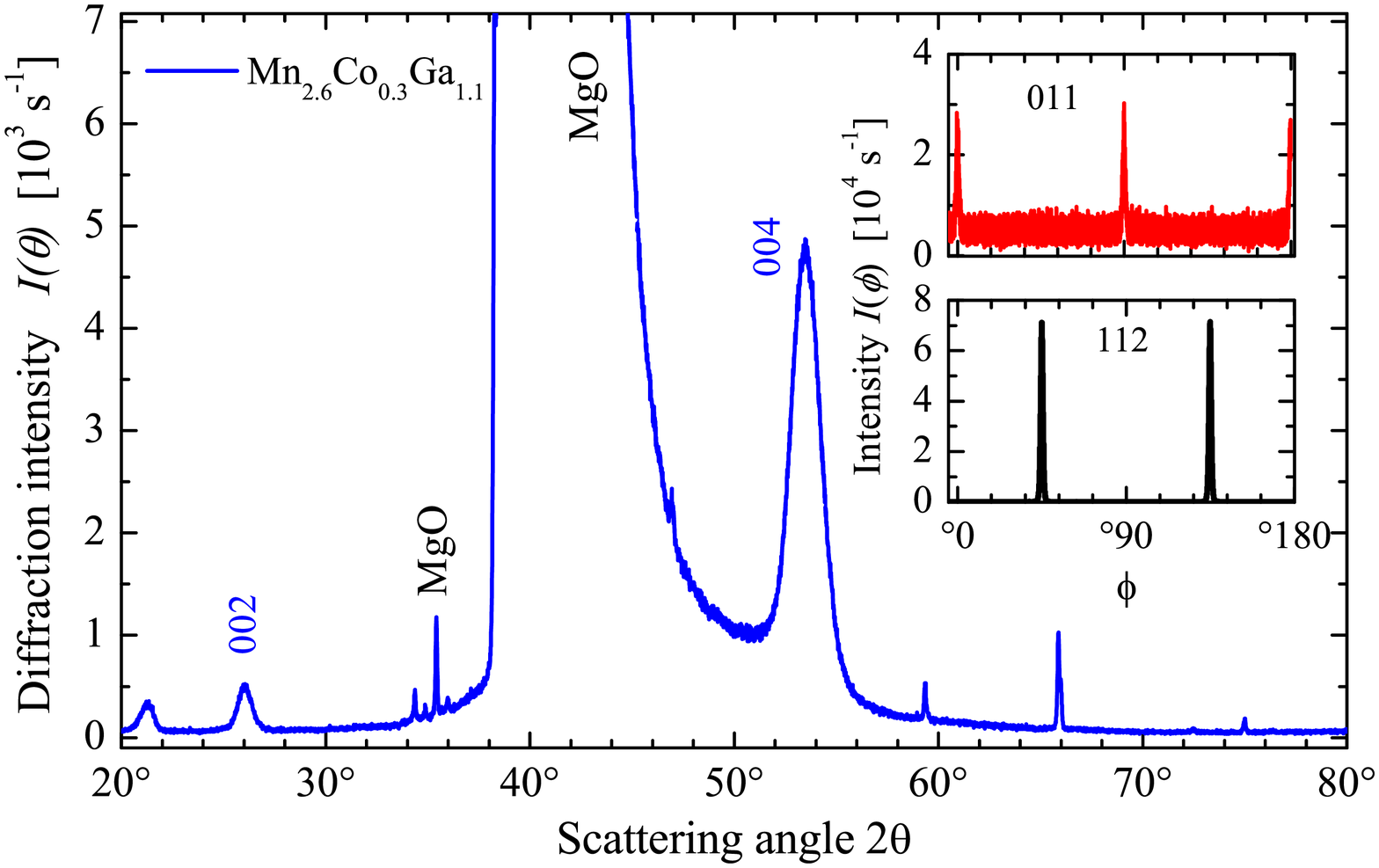}
  \caption{X-ray diffraction pattern for 30-nm-thick Mn$_{2.6}$Co$_{0.3}$Ga$_{1.1}$ film.
           The insets show the azimuthal scans of the (011) (top) and (112) (bottom) planes.}
\label{fig:XRD}
\end{figure}

Figure~\ref{fig:XRD} shows the XRD 2$\theta$-$\omega$ pattern of 
Mn$_{2.6}$Co$_{0.3}$Ga$_{1.1}$ film. In addition to the peaks originating due to 
MgO, only the (002) and (004) peacks can be observed, thereby indicating that the 
Mn$_{2.6}$Co$_{0.3}$Ga$_{1.1}$ film is grown with the tetragonal $c$ axis, wish is along 
the normal direction. The lattice constants are $a=3.94$~{\AA} and $c=6.85$~{\AA}. 
Furthermore, azimuthal ($\varphi$) scans were employed to confirm the crystalline 
structure of the film. The presence of (011) reflex in the film--as shown in 
the inset of Figure~\ref{fig:XRD}--indicates crystallization in a tetragonal 
D0$_{22}$-type structure~\cite{WMW09}. The Mn$_{2.1}$CoGa$_{0.9}$ film shows a cubic 
structure with a lattice constant of $a = 5.889$~{\AA}.
 
%%%%%%%%%%%%%%%%%%%%%%%%%%%%%%%%%%%%%%%%%%%%%%%%%%%%%%%%%%%%%%%%%%%%%%
%%%%%%%%%%%%%%%%%%%%%%%%%%%%%%%%%%%%%%%%%%%%%%%%%%%%%%%%%%%%%%%%%%%%%%

\begin{figure}[htb]
  \centering
  \includegraphics[width=9cm]{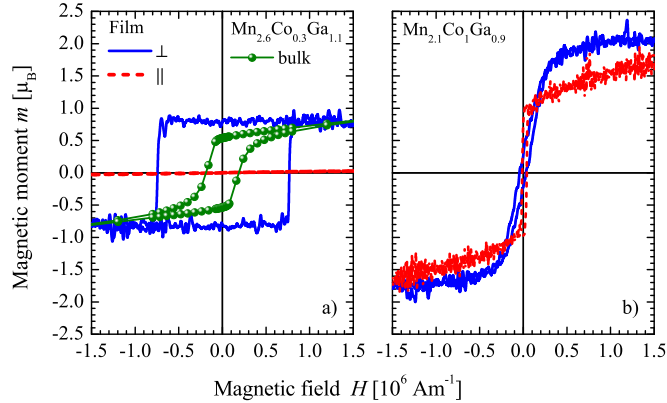}
  \caption{Perpendicular($\perp$) and in--plan($\parallel$) hysteresis curves of (a) tetragonal 
  Mn$_{2.6}$Co$_{0.3}$Ga$_{1.1}$ and (b) cubic Mn$_{2.1}$Co$_{1}$Ga$_{0.9}$ thin films.}
\label{fig:M}
\end{figure}

Figures~\ref{fig:M}(a) and ~\ref{fig:M}(b) show the hysteresis loops of the 
Mn$_{2.6}$Co$_{0.3}$Ga and Mn$_{2.1}$CoGa$_{0.9}$ films, respectively. The 
magnetic field was applied perpendicular ($\perp$) or in-plane ($\parallel$) to 
the film plane direction. When a magnetic field is applied perpendicular to the 
film plane, the magnetization curves of the film with less Co content 
(Mn$_{2.6}$Co$_{0.3}$Ga) exhibit a rectangular shape, whereas those of 
Mn$_{2.1}$CoGa$_{0.9}$ exhibit a soft magnetic behavior. 

It is significant that the coercitive field $H_c$ of the Mn$_{2.6}$Co$_{0.3}$Ga film is large when 
compared with the corresponding values reported previously~\cite{MKW12}. 
The results indicate that the easy axis of the 
magnetization is perpendicular to the film plane in the tetragonal case. 
For comparison, the hysteresis of polycrystalline Mn$_{2.6}$Co$_{0.3}$Ga bulk material was 
measured; the results are shown in Fig.~\ref{fig:M}(a). The coercive field is obviously smaller compared with that of the thin film.
The bulk sample is not saturated even at fields of $\mu_0H=6$~T (outside the range shown in the figure).
However, the magnetic moments per formula unit are of the same order as those of the films.
This non--saturating behavior is typical for polycrystalline samples consisting of randomly oriented grains with different alignment
of the easy or hard magnetization directions.

%%%%%%%%%%%%%%%%%%%%%%%%%%%%%%%%%%%%%%%%%%%%%%%%%%%%%%%%%%%%%%%%%%%%%%
%%%%%%%%%%%%%%%%%%%%%%%%%%%%%%%%%%%%%%%%%%%%%%%%%%%%%%%%%%%%%%%%%%%%%%

The uniaxial anisotropy constant K$_u$ was estimated using the relation as 
described in a previous work~\cite{MKW12}. The effective anisotropy field $H_{eff}$ was 
obtained by the extrapolated intersection of the in--plane $M-H$ curve with the 
saturation magnetization value of the perpendicular $M-H$ curve. The magnetic 
moment $m$, uniaxial magnetic anisotropy $K_u$, and coercive field H$_c$ of the 
films are compared in Table~\ref{tab:MKH}. The Co--doped film 
Mn$_{2.6}$Co$_{0.3}$Ga$_{1.1}$ shows a high uniaxial magnetic anisotropy K$_u$ as 
 previously reported magnetic anisotropy values for Mn-Ga films. Further, the film exhibits a high 
coercive field of $7.74\times10^5$~A~m$^{-1}$ and remanence of 0.33~T. These 
values are comparably higher than those obtained for polycrystalline bulk 
materials. The cubic Mn$_{2.1}$CoGa$_{0.9}$ shows a soft magnetic behavior, 
as reported for half metallic Heusler compounds~\cite{AWF11}.

%%%%%%%%%%%%%%%%%%%%%%%%%%%%%%%%%%%%%%%%%%%%%%%%%%%%%%%%%%%%%%%%%%%%%%
\begin{table*}[htp]
\centering
\caption{Composition dependence of magnetic moment $m$ (per formula unit), uniaxial 
         magnetic anisotropy K$_u$, coercive field $H_c$, remanence $B_r$, 
         specific maximum energy product $(BH)_{max}$, and specific energy integral $W_h$ 
         of different Mn$_{3-x}$Co$_{x}$Ga films compared with bulk materials. The bulk
         Mn$_{2.6}$Co$_{0.3}$Ga$_{1.1}$ was not saturated.}
  \begin{ruledtabular}
  \begin{tabular}{lcccccc}
                                          &   $m$    &     $K_u$    &    $H_c$     & $B_r$ & $(BH)_{max}$ \\
                                          & $\mu_B$  & M J m$^{-3}$ & k A m$^{-1}$ &   T   & k J m$^{-3}$ \\
      \hline
  Mn$_{2.1}$CoGa$_{0.9}$    (cubic)       &   2.03   &       -      &     24       & 0.08  & 0.38         \\
  Mn$_{2.6}$Co$_{0.3}$Ga$_{1.1}$          &   0.84   &     1.2      &    757       & 0.33  & 24           \\
  Mn$_{3}$Ga [Ref.~\cite{MKW12}]          &    -     &   1.0 - 1.5  &     -        &  -    & -            \\
  Mn$_{3}$Ga [Ref.~\cite{KRV11}]          &    -     &     0.89     &     -        &  -    &             \\
      \hline
  Mn$_{2.6}$Co$_{0.3}$Ga$_{1.1}$ bulk     &   n.s.   &       -      &    199       & 0.12  &  2.61        \\
  Mn$_3$Ga bulk [Ref.~\cite{BFW07,WBF08}] &   1.0    &       -      &    453       & 0.136 &  18.3        \\
  \end{tabular}
  \end{ruledtabular}
\label{tab:MKH}
\end{table*}
%%%%%%%%%%%%%%%%%%%%%%%%%%%%%%%%%%%%%%%%%%%%%%%%%%%%%%%%%%%%%%%%

The magnetic materials are also characterized by various magnetic energies 
besides anisotropy, coercive field, and remanence~\cite{Co10}. 
An important energy parameter of interest is the maximum energy product 
$BH_{max}=\max(-B\times H)$. It represents the maximum useful magnetic energy of 
a permanent magnet. Its value is obtained by multiplying $B$ times $H$ in the 
second (or fourth) quadrant of the hysteresis loop. The specific energy integral 
$W_H=\oint HdB$ is from direct integration of the magnetization loops 
and the hysteresis loss per cycle. For an ideal hard magnet, a nearly 
rectangular hysteresis loop ($B(H)$) with $W_H \approx 4 \times BH_{max}$ is 
expected. Magnetic energies of the different samples are compared in 
Table~\ref{tab:MKH} together with the other magnetic data.

\begin{figure}[htb]
\centering
\includegraphics[width=7cm]{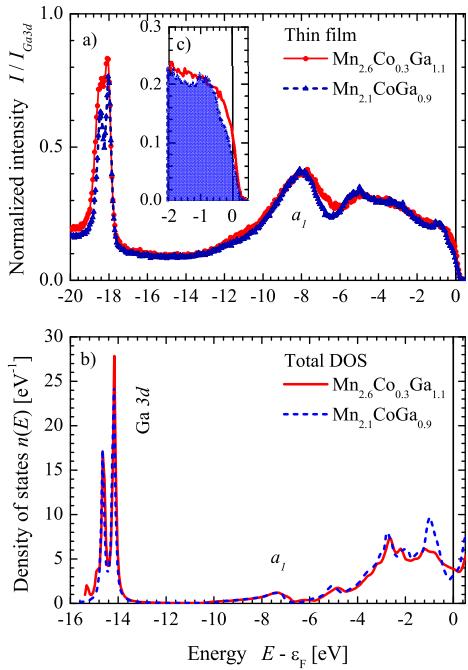}
\caption{(a) Valence band spectra and (b) total density of states of 
          Mn$_{2.6}$Co$_{0.3}$Ga$_{1.1}$ and Mn$_{2.1}$Co$_{1}$Ga$_{0.9}$ films. 
        (c) Region close to the Fermi energy $\epsilon_F$ on an enlarged scale.}
\label{fig:dosvb}
\end{figure}

The electronic structures of Mn$_{2.6}$Co$_{0.3}$Ga$_{1.1}$ and 
Mn$_{2.1}$Co$_{1}$Ga$_{0.9}$ were investigated using HAXPES at room temperature. 
Figure~\ref{fig:dosvb} compares the measured valence band spectra to the 
calculated total density of states (DOS). All the major structures observed in the 
spectra are in good agreement with the calculated DOS. The intensity ratios of the 
peaks are different from the calculated DOS. This deviation arises from the 
different partial cross sections of the $s$, $p$, and $d$ states that are localized at 
different atoms of the material~\cite{FBO07}.

The low--lying maximum at about -8~eV below the Fermi edge $\epsilon_F$ arises 
from the $a_1$ ($s$) states being localized at the Ga atoms. These states are broader for 
Mn$_{2.6}$Co$_{0.3}$Ga$_{1.1}$ with an additional shoulder being present at about -6.7~eV. 
Such states arise due to the presence of the additional Ga atoms, which are located at the 
$2b$ Wyckoff position that is normally occupied by Mn.

An interesting feature here is the significant changes observed in the spectra close to 
$\epsilon_F$, as shown in Figure~\ref{fig:dosvb}(c), where the spectral values indicate the 
difference between the tetragonal Mn$_{2.6}$Co$_{0.3}$Ga$_{1.1}$ and the cubic 
Mn$_{2.1}$CoGa$_{0.9}$ films. In the cubic case, a sharp peak at -0.94~eV 
is evident, whereas for the tetragonal films, the corresponding states are smeared out. This 
change in the electronic structure is due to a $van~Hove$ singularity occurring close 
below the Fermi energy. Changing the composition of the compound shifts this 
van~Hove singularity to the Fermi energy. This causes a {\it band
Jahn-Teller effect} that results in the tetragonal distortion of the crystalline 
structure upon changing the number of valence electrons~\cite{WCG12}.

%%%%%%%%%%%%%%%%%%%%%%%%%%%%%%%%%%%%%%%%%%%%%%%%%%%%%%%%%%%%%%%%%%%%%%
%%%%%%%%%%%%%%%%%%%%%%%%%%%%%%%%%%%%%%%%%%%%%%%%%%%%%%%%%%%%%%%%%%%%%%

An inspection of Figure~\ref{fig:dosvb} might lead to the assumption that the obtained data contradicts 
the above statement as both the DOS and the intensity at $\epsilon_F$ are seemingly higher
in the tetragonal case. Here, it should be mentioned that the high density in the cubic case is still tot close to
 $\epsilon_F$. Further, the molecular Jahn--Teller effect is not completely valid for solids.
Instead of observing a bare splitting of the states at the $\Gamma$-point, we can observe 
changes in all energy bands that are coupled to each other.
The energy gain by removing the van~Hove singularity from the Fermi energy is given by the change in the band energy
$E_{\rm Band}=\int_0^{\epsilon_F} E\cdot N(E)dE$ when the density of states $N(E)$ changes with the structure.
The obtained result is different from the energy gain obtained by the simple splitting of a state.

%\section{Summary and conclusions}
%%%%%%%%%%%%%%%%%%%%%%%%%%%%%%%%%%%%%%%%%%%%%%%%%%%%%%%%%%%

In summary, epitaxial thin films of Mn$_{3-x}$Co$_{x}$Ga were grown with 
varying levels of Co content. Depending on the Co content level, tetragonal and cubic 
structures were is performed. The composition dependence of the saturation magnetization 
$M_S$ and uniaxial magnetic anisotropy $K_u$ in the epitaxial films were 
investigated. A high magnetic anisotropy $K_u$ of 1.2~MJ m$^{-3}$ was achieved 
for the tetragonal Mn$_{2.6}$Co$_{0.3}$Ga$_{1.1}$ film with a low magnetic moment 
$m_S$ of 0.84$\mu_B$.

Furthermore, the valence band of the films was examined by HAXPES to study the 
structural dependence of the electronic structure of the Co--doped Mn-Ga films. A 
van Hove singularity close to the Fermi edge is observed in the cubic films, 
whereas the corresponding energy states are smeared out in the tetragonal films 
due to the presence of the band Jahn-Teller distortion.

%%%%%%%%%%%%%%%%%%%%%%%%%%%%%%%%%%%%%%%%%%%%%%%%%%%%%%%%%%%%%%%%%%%%%%
\bigskip 
\begin{acknowledgments}

The authors gratefully acknowledge financial support by the DfG-JST (P~1.3-A and 
2.1-A in FOR 1464 ASPIMATT). The synchrotron--based HAXPES measurements were 
performed at the beamline BL47XU of SPring-8 facility with the approval of JASRI (Proposal No.~2012A0043).

\end{acknowledgments}

%%%%%%%%%%%%%%%%%%%%%%%%%%%%%%%%%%%%%%%%%%%%%%%%%%%%%%%%%%%%%%%%%%%%%%
%\bibliography{MnCoGa_TF}


\end{document}